\documentclass[12pt]{article}
\input{epsf.tex}
\begin{document}
\begin{center}
{\bf SOME PROBLEMS IN DETERMINING LEVEL DENSITY AND RADIATIVE STRENGTH
FUNCTIONS IN LIGHT AND NEAR-MAGIC NUCLEI}
\end{center}
\begin{center} 
{\bf  A.M. Sukhovoj, V.A. Khitrov, Li Chol}\\
{\it FLNP, Joint Institute for NuclearResearch, Dubna, Russia}\end{center}
%\end{document}
\begin{center} 
{\bf Pham Dinh Khang}\\ {\it National University of Hanoi, Vietnam}\end{center}
\begin{center} 
{\bf Nguyen Xuan Hai,
Vuong Huu Tan}\\{\it
Vietnam Atomic Energy Commission, Vietnam }\\
\end{center}
\begin{abstract}
The values of some functional dependencies of level density and radiative
strength functions that reproduce the experimental intensities of the two-step
gamma-cascades to the ground and first excited states of $^{28}$Al have been
determined. It was shown that the assumption about independence of the
dipole cascade transitions radiative strength functions on energy of decaying
level leads to rather essential error in observation of both level density
and radiative strength functions.
\end{abstract}
\section*{Introduction} \hspace*{14pt}

High level density in most nuclei does not allow determination of their
parameters (spin, parity and so on) and, consequently, these data cannot
be involved in calculation. In these cases, there are usually used theoretical
notions on the level density $\rho$ and radiative strength functions $k$:
\begin{equation}
k=f/A^{2/3}=\Gamma_{\lambda l}/(E_\gamma^3\times A^{2/3}
\times D_\lambda).
\end{equation}
This definition of $k$ for used gamma-transition energy  $E_\gamma$ and
spacing  $D_\lambda$ between the decaying levels  $\lambda$ provides week
dependence on nuclear mass $A$ and allows direct comparison of $k$ for the
nuclei with sufficiently different masses. In this case it is implied that the
partial width $\Gamma_{\lambda l}$ of the transitions between the levels
$\lambda$ and  $l$ is averaged over some set $m=\rho \times \Delta E$ of its
random values from a given energy interval $\Delta E$.
But such approach is considered as inapplicable for light and near-magic
nuclei with low level density and clearly expressed influence of nuclear
structure on the gamma-transition spectrum. It seems useful to determine
experimentally the energy region where the notions of level density and
radiative strength functions can be used. Only this argument stipulated our
effort to determined the mentioned parameters for the $^{28}$Al compound
nucleus [1]. It should be noted, that up to now the reliable data on the
level density and radiative strength functions can be derived mainly from 
the experimental intensities of the two-step gamma transitions [2]: 
\begin{equation}
I_{\gamma\gamma}=\sum_{\lambda ,f}\sum_l\frac{\Gamma_{\lambda l}}{
\Gamma_\lambda}\frac{\Gamma_{lf}}{\Gamma_l}=\sum_{\lambda ,f}
\frac{\Gamma_{\lambda l}}{<\Gamma_{\lambda l}> m_{\lambda l}} n_{\lambda
l}\frac{\Gamma_{lf}}{<\Gamma_{lf}> m_{lf}},
\end{equation}
proceeding between the compound state (neutron resonance) and a group of
low-lying levels of the nucleus under study that were determined according to
algorithm [3] for all the possible energy intervals $\Delta E$ of their
primary gamma-transitions $E_1$.
Analogous data derived from the spectra of evaporated nucleons or gamma-ray
spectra following decay of levels $E_{ex}$ excited in nuclear reactions have too
large systematic errors. They are conditioned by the use of theoretical models
for calculating penetration coefficients of nuclear surface for evaporated
nucleons or by very large (at least several hundreds) transfer 
coefficients of experimental systematic errors into the determined 
$\rho$ and  $k$ values or by principal incorrectness  [4,5] of the data processing
procedure. In the first case, one cannot expect required accuracy in
determination of level density and its energy dependence below,
at least, neutron binding energy.
In the second case it is necessary to get realistic estimations of systematic
uncertainties and to reduce them to minimum possible magnitude using all the
possibilities of experiment, mathematic and mathematical statistics.

\section {Specific in determination of  $\rho$ and $k$ from experimental
spectra} \hspace*{14pt}
Equation (2) contains more parameters than the number of selected intervals
but the  range of possible variations for these parameters  is always
limited and rather narrow if:

(a) experimental spectra are decomposed into two components containing solely
primary or solely secondary gamma-transitions according to method [3] based
on some grounds; 

(b) the ratio  $\Gamma_{\lambda l}$ and $\Gamma_{lf}$ of partial widths for the
primary and secondary transitions of equal energy and multipolarity is set.

Minimization of discrepancy between the experimental and determined within
iterative process [2] $\rho$ and  $k$ requires also to use the density of
neutron resonances, number of low-lying levels and total radiative width of
$s$-resonances from [6]. Process [2] quickly enough converges to some average
value of the cascade gamma-decay parameters (in limits of principally
irremovable scatter) if experimental distribution (2) does not have
considerable local bumps. Acceptable values of the parameter $\chi^2$ for
$^{28}$Al are achieved sometimes only after 50000-100000 iterations.
In any case, no correlation between varying input parameters $\rho$ and $k$ and
deviation of their best values from the average was found.

As it is seen from Fig.~1, the mentioned above condition is not fulfilled
in case of $^{28}$Al and there are observed serious structure effects.
Moreover, these effects weakly enough manifest themselves in instrumental
spectrum and very strongly in  dependence of cascade intensity on energy of
the primary transition. This testifies to necessity to use method [3] for
extraction of $\rho$ and  $k$ from the experiment and shows uncertainty of
the results of analysis like that described in [7].

Usually model reproduction of the cascade gamma-decay process of high-lying
levels ($E_{ex} \approx B_n$)  is performed within hypothesis of independence
of radiative strength functions on the energy of decaying level. 
The values of $\rho$ and  $f=kA^{2/3}$ [8] were determined from the data on
the reactions $^{28}$Si($^3$He,$^3$He,$\gamma)^{28}$Si and  
$^{28}$Si($^3$He,$\alpha \gamma)^{27}$Si using this assumption.
This assumption cannot be used for the excitation energy, $E_{ex} <0.5 B_n$
even for nuclei with high level density what follows [9] from the experimental
and calculated population of these levels. The direct, although not exhaustive
notion about the function $k=F(E_\gamma, E_{ex})$ can be obtained from the
experimental data on the two-step gamma-cascades  only if there is observed
not less than 95-99\% of their intensity.

However, at present such statistics of information is unattainable.
Therefore, it is of practical interest even the data on the sign of deviation
of $\rho$ with respect to the experimental value when the standard approach
$k=F(E_\gamma ,E_{ex})/F(E_\gamma ,B_{n})=$const is used in analysis [2].

\subsection{Spectrum of possible functions $\rho$, $k$ and their most
probable values}\hspace*{14pt}

Fig.~1 demonstrates intensity spectrum obtained from experimental spectra by
re-distribution of cascade intensities resolved as the pair of peaks in so
that their intensity is summed in corresponding energy intervals of the
primary transitions (and subtracted  from energy interval of the secondary
transitions). Low level density in $^{28}$Al results in absence of continuous
distribution of a number low intensity cascades. Combination of the method [10]
to determine quanta ordering in cascade with the data [11,12] from the
evaluated decay scheme provides extremely small systematic errors in both
total cascade intensity and its energy dependence.
In further analysis we neglect them. Sufficiently larger errors in the
observed $\rho$ are caused by the use of the assumption
$F(E_\gamma, B_{n})/F(E_\gamma, E_{ex})=$ const.

But their magnitude can be estimated only in indirect way. Fig.~2 shows the
results of determination of the cascade gamma-decay parameters for three
variants of analysis:

(a) the level density is fixed and set by the non-interacting Fermi-gas model;

(b) the level density and radiative strength functions are varied parameters
of the process [2];

(c) the function $\rho$ below $E_{ex}\approx $ 7 MeV is set by the simplest
interpolation of the cascade intermediate levels observed in [1].

Variants (a) and (c) do not provide precise reproduction of $I_{\gamma\gamma}$
in principle. This caused with anomalous high experimental values of $k$ for
the primary $E1$ transitions with the energy  $E_1 \le 3.5$  MeV and
very small for the energy exceeding 4.5 MeV (multipolarity of the primary and
secondary transitions in this nucleus is equal due to corresponding parity of
corresponding levels). This can be compensated  by extremely high and
absolutely unreal level densities at $E_{ex} >4$ MeV, as it follows from the
variant (b). But observed in this case radiative strength functions of the
cascade primary transitions are several order of magnitude less than the
values of $k$ derived from known [10] primary transition intensities according
to eq.(1). The Porter-Thomas fluctuations of such scale are to small in order
to explain the data presented in Fig.~2. It should be noted here that all
the known levels with negative parity from the energy interval
$3.7 < E_{ex} <5.7$ MeV are populated by intense primary transitions and this
effect cannot be explained by random fluctuations.
Accounting for the fact that more than 90\% of the experimental spectra area
[1] of the nucleus under consideration is concentrated as resolved peaks,
one can assume that the main portion of the desired level density and
radiative strength functions is caused just by individual cascades.
The last variant of method [2] takes into account possible inequality of
level density with different parities. Therefore, impossibility of precise
reproduction of $I_{\gamma\gamma}^{exp}$ using level densities (a) and (c) have
to connect only with change in energy dependence of $k$ when changing the
energy of decaying level.

Unfortunately, practically there is no possibility to estimate the value of
this effect within the method [9] for  $^{28}$Al owing to large difference
in energy dependence of radiative strength functions for primary and
secondary transitions and, correspondingly, very strong change in the
determined $\rho$ values. In heavier nuclei, corrections of function
$k=F(E_\gamma, B_{n})$ required to derive $F(E_\gamma, E_{ex})$ change 
level density  sufficiently weaker.

Possibility of the practically precise reproduction of experimental
distribution $I_{\gamma \gamma}$ by values $\rho$ and $k$ with the large
systematic errors on the condition:
$\sum (dI/d \rho \times \delta \rho) = - \sum (dI/dk \times \delta k)$ 
it leads to the appearance in the procedure [2] of the false solutions.
Mathematically this is equivalent to the presence in function of the maximum
likelihood of false maximums. Mathematical statistics provides for this
possibility and recommends for their identification the using of a
maximum spectrum of the initial values of the parameters of iterative process.

With the multiple repetition of random process [2] the probability of equality
$\sum (dI/d\rho\times \delta \rho)=-\sum (dI/dk \times \delta k)$ it diminishes
with an increase in the sum
$\sum (|dI/d\rho\times \delta \rho|+|dI/dk \times \delta k |)$.
Therefore the average value of the obtained parameters  will be most probable.
Of course, this statement is true only if relation (2) between the measured functions
and desired parameters corresponds to reality. Its discrepancy can be discovered only
in the analysis of additional experimental information.
For example, there can be use of theoretical models that limit the region of possible
values of required parameters. The results shown in Fig.~2 clearly illustrate this
situation because in the frameworks of the level density exceeding of this parameter
over the predictions of the non-interacting Fermi-gas model is impossible.
Similar problem arises also when  $\rho$ and  $k$ are extracted from the total gamma-ray
spectra depopulating the levels excited in reaction like $(^3$He,$^3$He$\gamma)$.
Possible scale of the problem in this case can significantly exceed that for two-step
cascades because intensity of the two-step cascades directly depends on the absolute value
of level density. But the intensity of primary gamma-transitions, extracted from the
nuclear reactions gamma-spectra  is not depend on the absolute value $\rho$.

\section {Full gamma-spectrum of the capture of the thermal neurons}

Comparison between the experimental [13] and calculated total gamma-spectra following
thermal neutron capture provides complementary information on confidence level and values
of possible systematical errors of the observed  $\rho$ and $k$. Results of this analysis
are shown in Fig.~3.

Theoretical and experimental spectra do not agree in all presented
variants of calculation. This means, first of all, that the desired parameters in each
variant a priory contain larger or smaller systematical errors. At every decay, the
considered nucleus gives of quite certain energy $B_n=7.725$ MeV independently on number
and energy of emitted quanta. This fact was used in [13] for normalization of the spectra
and its use in the experiment like [8] would provide serious decrease in systematic errors
of the determined there $\rho$ and $k$ values. In analysis [3], however, this circumstance
is, most probably, negative: large variations of calculation parameters are reduced in
essentially less errors of calculated spectra. This means that relatively small discrepancy
between calculation and experiment can be related with large errors of $\rho$ and $k$.
 
In all variants, calculated intensity of low-energy ($E_\gamma, <2$ MeV) gamma-quanta is
noticeably less than that experimental. This can be caused either by systematic errors in
data [13] or by large values of radiative strength functions of the secondary
gamma-transitions in mentioned region of their energy. Maximum exceeding of theoretical
intensity in variant (b) results, most probably from excessive value of level density in
Fig.~2b. The least discrepancy of theoretical and experimental intensities above 7 MeV
can be achieved only if the total calculated intensity of all primary transitions of
smaller energy has minimal value (most probably due to the least systematic errors of level
density in this set of considered $\rho$ and $k$.

\section{Conclusion}
Data analysis about the two-step cascades, obtained during capture of thermal neutrons
into $^{27}$Al, revealed the strong divergence of the form of the energy dependence 
of radiative strength functions for the gamma-transitions of one and the same energy and
multipolarity during decaying of their initial and intermediate level respectively.
Effect has considerably larger value, than  this is observed in the more heavy nuclei [9].
General trend of energy dependence of the sums of radiative strength functions providing
the best reproduction of the experimental intensity of the two-step cascades corresponds to
the data obtained from the absolute intensities of the primary gamma-transitions.
It demonstrates increase in strength functions of probably low-energy $M1$ transitions
depopulating the compound state.
This conclusion is correct if the density of the cascade intermediate levels unessentially
differs from the simplest extrapolations of exponential type.\\ 

\newpage
{\bf REFERENCES }\\\\
1. Bondarenko V.A., Honzatko J., Khitrov V.A., Sukhovoj A.M., \\\hspace*{14pt}
Tomandl I.
  Two-step cascades following thermal neutron capture in $^{27}$Al,\\\hspace*{14pt}
  Fizika B (Zagreb), 2003, {\bf 12(4)}, 299.\\
2. Vasilieva E.V., Sukhovoj A.M., Khitrov V.A.,
Direct Experimental \\\hspace*{14pt}Estimate
of Parameters That Determine the Cascade Gamma
Decay of\\\hspace*{14pt Compound States } of Heavy Nuclei,
Phys. At. Nucl. 2001, {\bf 64(4)}, 153.
\\ \hspace*{14pt}  Khitrov V. A., Sukhovoj A. M.,
New technique for a simultaneous \\\hspace*{14pt estimation of the} level density and
radiative strength functions of
dipole\\\hspace*{14pt} transitions at $E_ex< B_n -0.5$ MeV //
INDC(CCP)-435, Vienna, 2002, 21\\
\hspace*{14pt} http://arXiv.org/abs/nucl-ex/0110017.\\
3. Boneva S. T., Khitrov V.A., Sukhovoj A.M.,
Excitation study of high-lying\\\hspace*{14pt states of } differently
shaped heavy nuclie by the method of \\\hspace*{14pt} two-step cascades
Nucl. Phys. 1995, {\bf A589}, 293.\\
4. Sukhovoj A.M., Khitrov V.A., Li Chol, On correctness of some processing  \\
\hspace*{14pt}operations for two-step cascade intensities data from the $(n_{th},2\gamma)$
reaction,\\\hspace*{14pt} JINR E3-2004-100, Dubna, 2004.\\\hspace*{14pt}
http://arXiv.org/abs/nucl-ex/0409016.\\
5. Khitrov V.A., Sukhovoj A.M., Pham Dinh Khang, Vuong Huu Tan,\\
\hspace*{14pt} Nguyen Xuan Hai, On the role of some sources of systematic errors
in\\\hspace*{14pt determination }of level density and radiative strength
functions from the\\\hspace*{14pt} gamma-spectra of nuclear reactions,
In: XI International Seminar on\\\hspace*{14pt} Interaction of Neutrons with Nuclei,
Dubna, 22-25 May 2003, E3-2004-9, Dubna, 2004,\\\hspace*{14pt} p. 107.\\
\hspace*{14pt}http://arXiv.org/abs/nucl-ex/nucl-ex/0305006.\\
6. Neutron Cross Section, vol. 1, part A, edited by Mughabhab S. F.,\\\hspace*{14pt} Divideenam M.,
Holden N. E., Academic Press 1981.\\
7. Be\v{c}v\'{a}\v{r} F. \sl et al.\rm, Phys.\ Rev.\ C \bf 52\rm,
(1995) 1278.\\\hspace*{14pt}
 Voinov A. \sl et al.\rm, Nucl.\ Instrum.\ Methods Phys.\ Res.\ A
\bf 497\rm, 350 (2003).\\
8. Guttormsen M. at all, J. Phys. G: Nucl. Part. Phys. {\bf 29} (2002) 263.\\
9. Bondarenko V.A., Honzatko J., Khitrov V.A., Li Chol, Loginov Yu.E., \\
\hspace*{14pt}Malyutenkova S.Eh., Sukhovoj A.M., Tomandl I.,
Cascade population of levels and \\\hspace*{14pt}probable phase transition in
vicinity of the excitation energy ~0.5Bn of heavy nucleus\\\hspace*{14pt}
In: XII International Seminar on Interaction of Neutrons with Nuclei, \\
\hspace*{14pt} Dubna, May 2004, E3-2004-169, p. 38.\\
 \hspace*{14pt}http://arXiv.org/abs/nucl-ex/0406030\\\hspace*{14pt}
http://arXiv.org/abs/nucl-ex/0410015 \\
10. Popov Yu.P., Sukhovoj A.M., Khitrov V.A., Yazvitsky Yu.S.,\\
\hspace*{14pt} Izv. AN SSSR, Ser. Fiz. {\bf 48} (1984) 1830.\\
11. http://www.nndc.bnl.gov/nndc/ensdf.\\
12. http://www-nds.iaea.org/pgaa/egaf.html.\\
13. Groshev L.V. et al. Atlac thermal neutron capture gamma-rays spectra,
\\\hspace*{14pt} Moscow, 1958.\\

\newpage
\begin{figure}%[htbp]
\vspace{-2cm}
\leavevmode
%\hspace-.8cm
\epsfxsize=13.0cm
\epsfbox{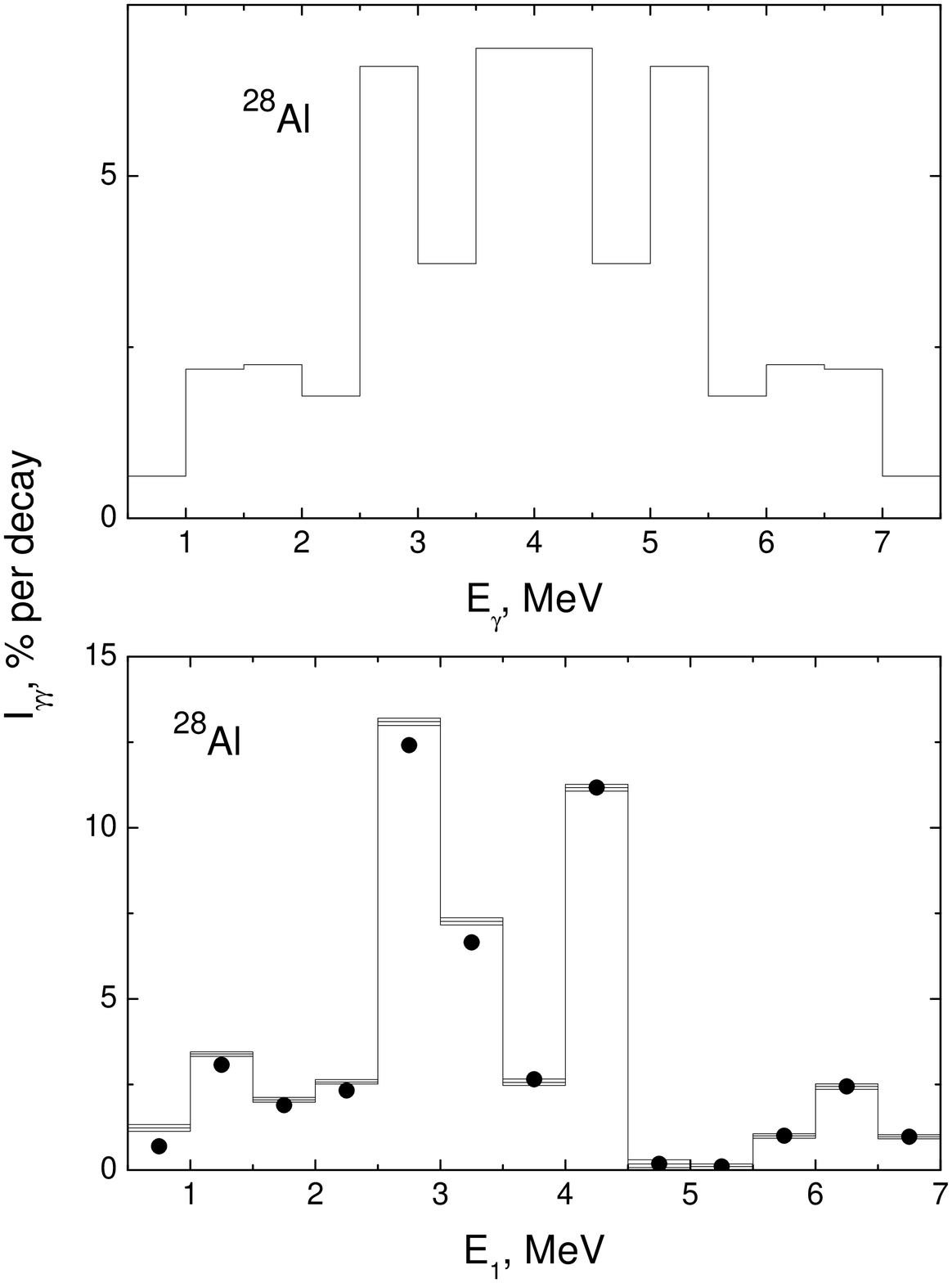}
%\vspace1cm

{\bf Fig. 1.} {\it Top: histogram represents intensities of two-step cascades in
the function of energy of primary, or second of gamma-transitions.
 Bottom:
histogram is the intensity of two-step cascades into $^{28}$Al,
 summed up [1] on interval of 0.5 MeV of their primary gamma-transition  energy only.
 Points - the same for the cascades, the resolved in the form of pairs peaks.}
\end{figure}
\begin{figure}%[htbp]\vspace-2cm
\leavevmode%\hspace-.8cm
\epsfxsize=13.0cm
\epsfbox {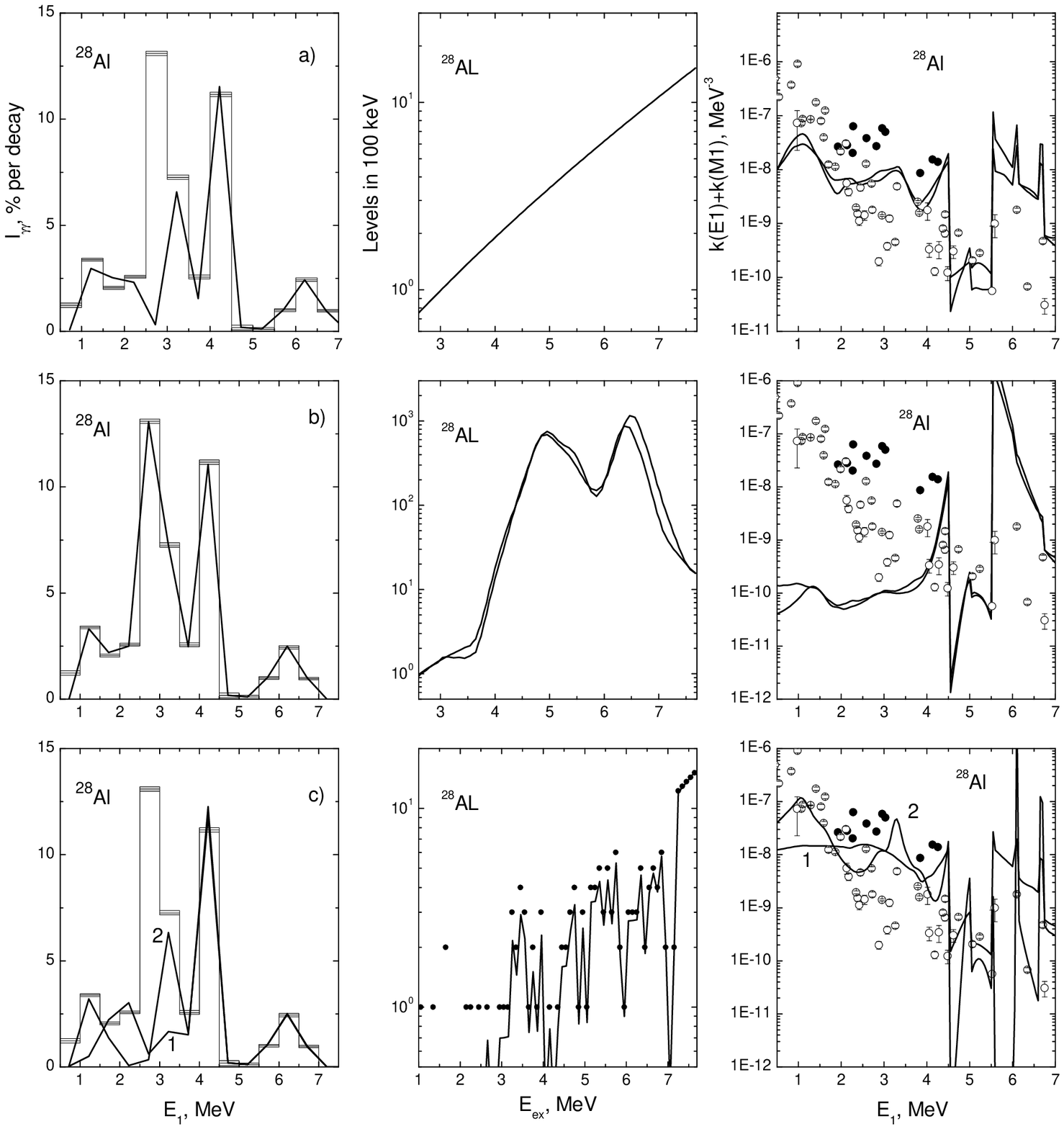}
\vspace{-6cm}

\bf Fig. 2. {\it Rows a)-c) are the versions of possible values $\rho$ and $k$,
maximally accurately
reproducing the intensity of two step cascades. Left column: histogram is experiment,
broken lines - best fit of the obtained values.
Points on the middle  column - number of obtained [1]  intermediate cascade levels.
Full points on by right column show the $k(E1)$ for the experimentally resolved primary
gamma-transitions
to the levels of negative [11,12] parity, the open points are the same for the levels
of the positive and
unknown parity.} \\
\end{figure}
\begin{figure}%[htbp]\vspace{-2cm}
\leavevmode%\hspace-.8cm
\epsfxsize=13.0cm
\epsfbox {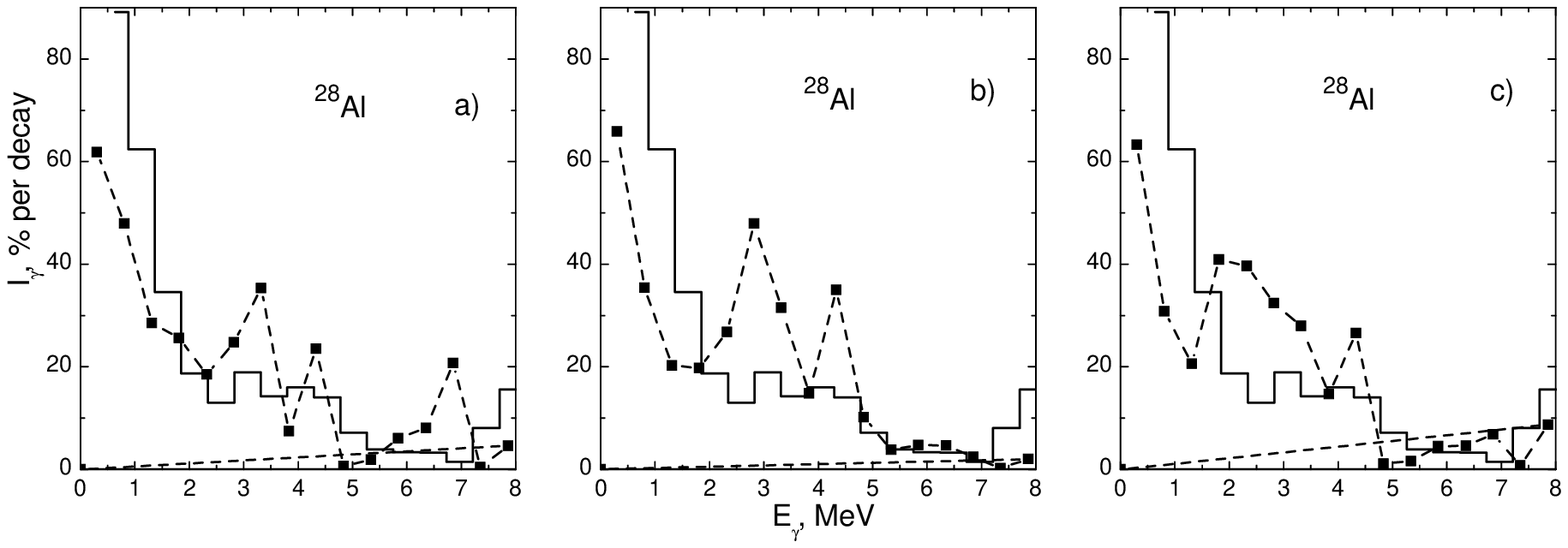}
\vspace{-13cm}

\bf Fig. 3. {\it ~The histogram represents the experimental gamma-quantum spectrum
following thermal neutron capture in aluminum. Points are its calculated values for
the density of levels
and radiative strength functions, represented in fig. 2a-2c.}
\end{figure}
\end{document}